\pdfoutput=1

\documentclass[11pt]{article}

\usepackage{ACL2023}
\usepackage{hyperref}
\usepackage{times}
\usepackage{latexsym}
\usepackage[T1]{fontenc}
\usepackage[utf8]{inputenc}
\usepackage{microtype}
\usepackage{inconsolata}
\usepackage{caption} 
\usepackage{subcaption}
\usepackage{multirow}
\usepackage{booktabs} 
\usepackage{amsmath,bm}
\usepackage{newfloat}
\usepackage{arydshln}
\usepackage{graphicx}
\usepackage{enumitem}
\usepackage{xspace}
\usepackage{float}
\newcommand{\tool}{{CommitBART}\xspace}

\begin{document}
\title{CommitBART: A Large Pre-trained Model for GitHub Commits}


\author{Shangqing Liu\textsuperscript{1}\footnotemark[1], Yanzhou Li\textsuperscript{1}\footnotemark[1], Xiaofei Xie\textsuperscript{2}, and Yang Liu\textsuperscript{1}\\ 
        \textsuperscript{1}Nanyang Technological University \\ \textsuperscript{2}Singapore Management University \\
        \{shangqin001, yanzhou001\}@e.ntu.edu.sg, xiaofei.xfxie@gmail.com, yangliu@ntu.edu.sg \\}
\maketitle
\def\thefootnote{*}\footnotetext[1]{Equal contribution.}

\begin{abstract}
GitHub commits, which record the code changes with natural language messages for description, play a critical role for software developers to comprehend the software evolution. 
To promote the development of the open-source software community, we collect a commit benchmark including over \textbf{7.99} million commits across \textbf{7} programming languages. Based on this benchmark, we present \tool, a large pre-trained encoder-decoder Transformer model for GitHub commits. The model is pre-trained by three categories (i.e., denoising objectives, cross-modal generation and contrastive learning) for six pre-training tasks to learn commit fragment representations. Furthermore, we unify a ``commit intelligence'' framework with one understanding task and three generation tasks for commits. The comprehensive experiments on these tasks demonstrate that \tool significantly outperforms previous pre-trained works for code. Further analysis also reveals each pre-training task enhances the model performance. 
\end{abstract}

\section{Introduction}
Large pre-trained models such as BERT~\cite{DBLP:conf/naacl/DevlinCLT19}, GPT~\cite{radford2019language} and T5~\cite{raffel2020exploring} have significantly improved state-of-the-art across a variety of natural language processing (NLP) tasks. These models are pre-trained on a large scale of unlabeled data with self-supervised objectives to learn contextual representations and then fine-tuned to multiple downstream tasks. Inspired by the great success of these models in NLP, recently a number of pre-trained models~\cite{feng2020codebert, ahmad2021unified, wang2021codet5} for programming languages (PL) have emerged in software engineering to advance the development of code intelligence. However, these works in the program scenario aim at learning the general program representation for a single function. The booming development of the open-source software industry has led to an unprecedented amount of projects hosted on Github\footnote{\url{https://github.blog/2018-11-08-100m-repos}}, which produces an extremely massive amount of commits. GitHub commits, which record the changed code (i.e., code changes) with the commit messages in natural language to describe the changed code, play a critical role for software developers to comprehend the evolution of software features at different stages of software development. To promote the development of the open-source software community, a specialized pre-trained model for ``commit intelligence'' is vital and meaningful for software developers.

The previous work CommitBERT~\cite{jung2021commitbert} proposed to directly employ CodeBERT~\cite{feng2020codebert} to fine-tune a model for commit message generation~\cite{jiang2017automatically, liu2020atom} (a task to automatically generate the natural language description given the input of the code changes) on the collected commit dataset (345K in total). 
However, CommitBERT did not pre-train a commit-specific pre-trained model on the commit data and this limits the performance on commit-related downstream tasks. Furthermore, the amount of the released data is far from satisfactory to the requirements for pre-training. 
In this work, we present \tool, a pre-trained encoder-decoder model based on BART~\cite{lewis2019bart} architecture for commits to support both commit-related understanding tasks and generation tasks. Compared with current pre-trained models for code, we are the first to provide a large pre-trained model for commits and unify a ``commit intelligence'' framework to support multiple commit-related tasks. 

Due to the lack of a large-scale commit benchmark, we collect a commit benchmark across \textbf{7} programming languages (i.e., C, CSharp, Java, JavaScript, PHP, Python and Typescript). Specifically, we extract commits from the top 500 projects based on the star ranking for each programming language to construct this benchmark. We obtain over \textbf{7.99} million commit data in total and make it public to benefit the follow-up researchers. We pre-train \tool using six pre-training tasks that can be divided into three categories. Specifically, they can be classified into denoising objectives (text infilling and graph-guided token masking), cross-modal generation (PL2NL generation and PLNL2PL generation) and contrastive learning (NLPL alignment and SimCSE~\cite{gao2021simcse}).
Besides text infilling~\cite{lewis2019bart}, to enhance semantic mapping between commit message and changed code, we design the objective of graph-guided token masking. It is designed to encourage the model to predict these tokens that both appear in the commit message and changed code. Furthermore, we design two cross-modal generation tasks: PL2NL generation task, which takes the changed code as input to generate its commit message, and PLNL2PL generation task, which takes the previous code as well as the commit message as the input to generate the updated code snippet. Both cross-modal generation tasks ensure that the model can generate high-quality code or natural language text about commits. We also design two contrastive learning-based objectives to enhance the semantics of code embeddings and message embeddings. The first NLPL alignment encourages the embedding of the changed code produced by the encoder to be closer to its corresponding message embedding and the second SimCSE takes the same input sequence to encoder twice with different dropout masks to aggregate semantic equivalent inputs.

We evaluate \tool with two widely concerned tasks for commits: security patch identification (understanding task) and commit message generation (generation task). We further propose two new commit-related generation tasks (i.e., positive code statements generation and updated code snippet generation). The extensive experiments illustrate that \tool achieves state-of-the-art performance on these tasks. Further analysis also reveals that each pre-training task enhances \tool to obtain better performance. The main contributions are summarized as follows:
\begin{itemize}[leftmargin=*]
    \item We collect a large-scale commit benchmark (over \textbf{7.99} million commits) across \textbf{7} programming languages and make it public for follow-up researchers.
    \item We are the first to present a large pre-trained model namely \tool for GitHub commits and it is pre-trained by the designed three categories for six pre-training tasks. 
    \item We unify a ``commit intelligence'' framework\footnote{All the code and data are available at \url{https://anonymous.4open.science/r/CommitBart-443E}} with one understanding task and three generation tasks for commits. The extensive experimental results on these tasks have demonstrated that \tool achieves state-of-the-art performance against the baselines. We encourage the follow-up researchers to contribute more commit-related tasks to this framework
\end{itemize}

\begin{table*}
\centering
  \caption{The statistics of our collected commit benchmark.}
  \label{tab:data-statistics}
  \scriptsize {\addtolength{\tabcolsep}{0pt}
  \begin{tabular}{crrrrrrrr}
    \hline
   Benchmark &C & CSharp & Java & JavaScript & PHP & Python & Typescript & Total\\
    \hline
    Pre-train & 1,917,109& 660,587 & 935,151 & 986,669 & 1,148,074 & 1,029,676& 762,760 & 7,440,026\\
    Fine-tune & 71,924 & 61,902 & 81,126 & 87,064 & 99,230 & 89,502& 67,762 & 558,510\\
    Total     & 1,989,033 & 722,489 & 1,016,277 & 1,073,733 & 1,247,304 & 111,9178 & 830,522 & 7,998,536 \\
    \hline
  \end{tabular}
  }
  \vspace{-4mm}
\end{table*}
\section{Background}
\subsection{GitHub Commits}
\begin{figure}[tp]
     \centering
     \includegraphics[width=0.99\columnwidth]{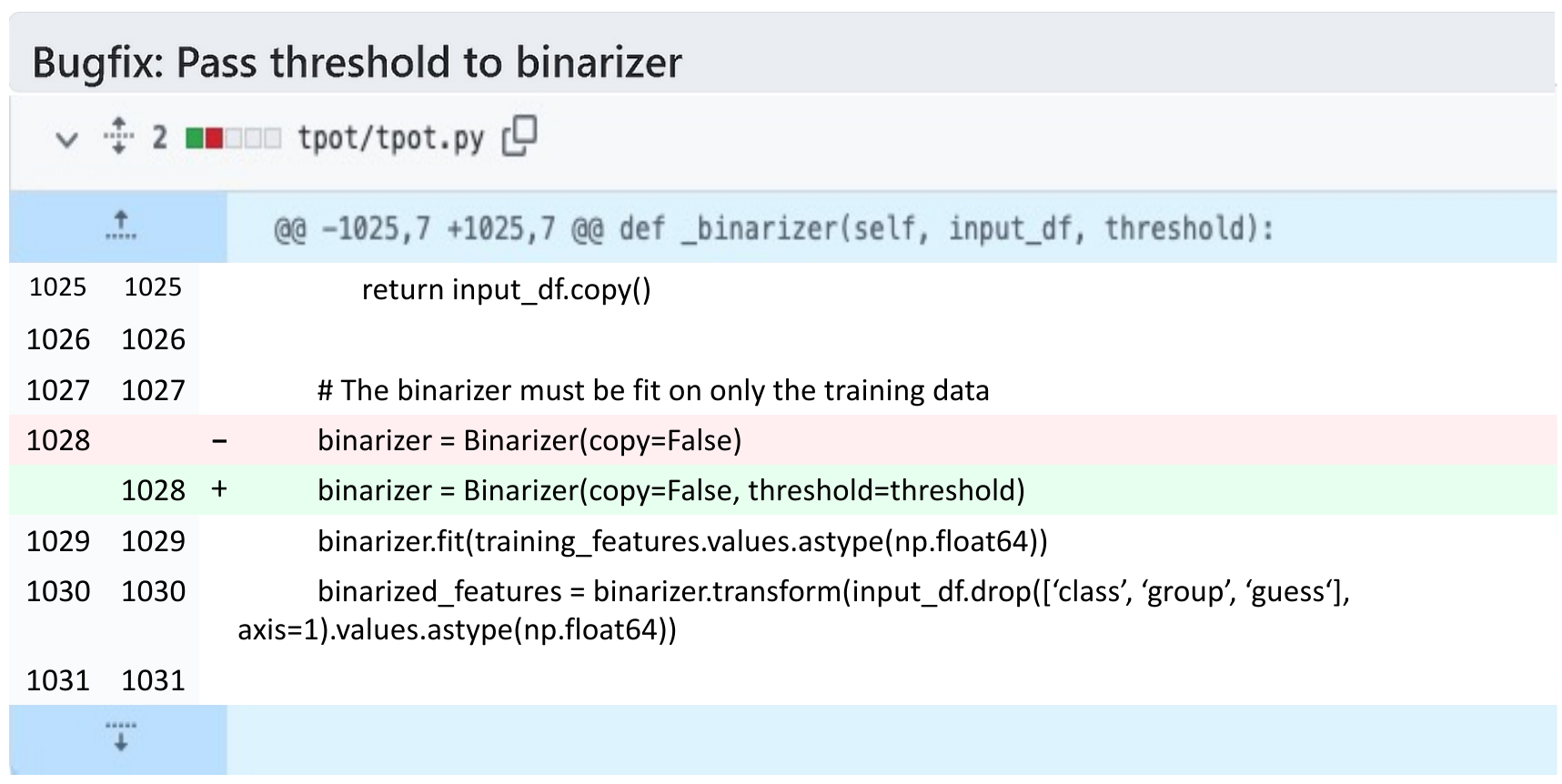}
    \caption{A commit with its commit id \href{https://github.com/rsumner33/tpot/commit/dbec56b8f813733bf24e9947747a242af3bd7d14}{dbec56}.}
     \label{fig:commit_example}
     \vspace{-6mm}
\end{figure}

A commit usually consists of the changed code with its commit message to describe the purpose of the current changed code in natural language. We present a commit to illustrate each component in Figure~\ref{fig:commit_example}. The upper rectangle contains the content of the commit message, which summarizes this commit in natural language (e.g., ``Bugfix: Pass threshold to binarizer''). The lower rectangle is the changed code namely one chunk, which contains the file path (i.e., ``tpot/tpot.py'') and the modified code from line 1025 to line 1031. The first line in this chunk starting with ``@@'' consists of the start line number (i.e., 1025) in the file, the total line statements (i.e., 7) and the function name (i.e., ``\_binarizer''). The changed code in this chunk is marked with ``+'' in the updated version of the code. Its previous version is marked with ``-''. The remaining content is the context around the changed code to reveal its contextual information. Hence, in summary, for this commit in Figure~\ref{fig:commit_example}, we can obtain that there is one line of code at line 1028 in the file of ``tpot.py'' changing the statement from ``...copy=False)'' to ``...copy=False, threshold=threshold)''. The other statements (i.e. from line 1025 to 1027 and from line 1029 to 1031) are the context of the statement at line 1028. Note that we just utilize a simple commit, which only has one chunk with single-line statement modification as the example for better illustration. In some cases, a commit may only have the updated statements or the deleted statements. They may not appear in pairs. Furthermore, a commit can change multiple places of code in a chunk or even change multiple files. Since commits play a critical role in the software development cycle, many researchers conduct broad research such as commit message generation~\cite{jiang2017automatically, liu2020atom, jung2021commitbert}, security patch identification~\cite{zhou2021spi, zhou2021finding, wu2022enhancing}. Compared with these existing works, which design a specialized neural model for the commit-related task, we are the first to propose a pre-trained model based on our large-scale benchmark and further unify these tasks with one model.

\subsection{Pre-trained Models in Code}
Pre-trained models for code have attracted widespread attention. For example, Svyatkovskiy et al.~\cite{DBLP:conf/sigsoft/SvyatkovskiyDFS20} proposed GPT-C, which utilizes the Transformer decoder~\cite{DBLP:conf/nips/VaswaniSPUJGKP17} in a left-to-right manner for code completion. CodeBERT~\cite{feng2020codebert} and CuBERT~\cite{DBLP:journals/corr/abs-2001-00059} pre-trained a bidirectional Transformer encoder on source code for some code-related understanding tasks. To further improve the model learning capacity, PLBART~\cite{ahmad2021unified} and CodeT5~\cite{wang2021codet5} are proposed to pre-train a Transformer encoder-decoder model based on BART and T5 to support both the understanding tasks and generation tasks. In addition, Guo et al.~\cite{guo2022unixcoder} proposed UniXcoder, a unified cross-modal pre-trained model based on a multi-layer Transformer for ``code intelligence''. Specifically, it utilized mask attention matrices with prefix adapters for code understanding and generation. Apart from the aforementioned works, there are also some pre-trained works for code~\cite{guo2020graphcodebert, DBLP:conf/uai/JiangZLLL21, DBLP:conf/nips/LuGRHSBCDJTLZSZ21, wang2021syncobert, buratti2020exploring, niu2022spt}. Compared with these pre-trained models for code, in this paper, we propose a large pre-trained model based on BART for GitHub commits and further unify a ``commit intelligence'' framework with different commit-related tasks.

\section{A Large-scale Benchmark for Commits}
Different from automated code review~\cite{li2022automating, tufano2022using, tufano2021towards}, which aims to review the code quality from GitHub pull requests, we target at GitHub commits and propose a large pre-trained model for commits to support commit-related intelligent tasks. The existing benchmark from CommitBERT~\cite{jung2021commitbert} only contains 345K commits which are far from the requirements for pre-training. To address this limitation, we collect a large-scale benchmark for commits instead. Specifically, we collect commits from the open-source projects on GitHub across \textbf{7} programming languages (i.e., C, CSharp, Java, JavaScript, PHP, Python, Typescript). To ensure the quality of the collected commits, we only keep the project whose description is in English and further select the top 500 projects based on their star ranking from January 2010 to December 2021 for each programming language through GitHub API~\footnote{\url{https://docs.github.com/en/rest}}. Given a cloned project, we utilize the open-source tool GitPython~\footnote{\url{https://gitpython.readthedocs.io}} to obtain the raw commits. These commits where the commit messages are non-English or the changed code length is greater than 2,000 are removed to reduce the sequence length. Furthermore, to ensure the quality of the commit message, followed by existing works~\cite{jiang2017automatically, liu2020atom, jung2021commitbert}, we only fetch the first sentence in the commit message as the target and filter out these commits where the target length less than three. To alleviate the learning process of the model overfit to the duplicated samples~\cite{allamanis2019adverse}, we conduct a strict de-duplication process to remove the same samples. Finally, we obtain over \textbf{7.99} million commits over \textbf{7} programming languages. For these \textbf{7.99} million commits, we further split them into pre-training data and fine-tuning data based on the ``project''. We just select those projects which have less than 2,000 commits to construct the fine-tuning dataset. The detailed statistics of our collected benchmark are presented in Table~\ref{tab:data-statistics}.

\section{CommitBART}
\begin{figure*}[tp]
     \centering
     \includegraphics[width=\textwidth]{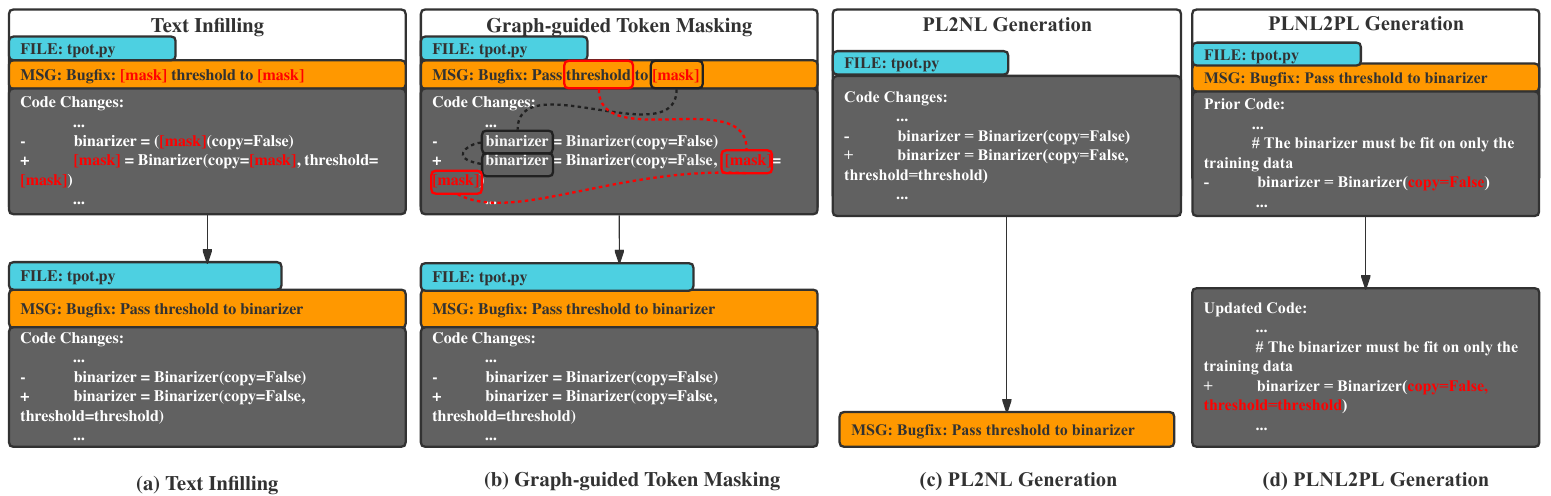}
    \caption{Denoising and cross-modal generation tasks for \tool.}
    \label{fig:four_tasks}
    \vspace{-6mm}
\end{figure*}

\begin{figure}[tp]
     \centering
     \includegraphics[width=\columnwidth]{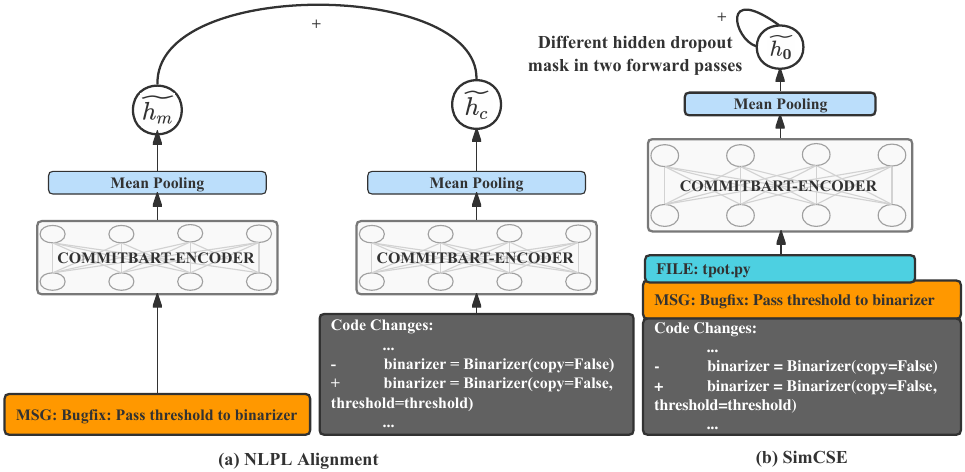}
    \caption{Contrastive pre-training tasks for \tool.}
     \label{fig:contrastive_tasks}
     \vspace{-6mm}
\end{figure}

\subsection{Model Architecture}
\tool follows the architecture of BART~\cite{lewis2019bart} and we adopt the parameters of PLBART~\cite{ahmad2021unified} to accelerate the training process. Since a commit usually consists of multiple different components such as the commit message, the updated code statements marked with ``+'', the deleted code statements marked with ``-'' and their context statements. To distinguish each component, we introduce some additional segment identifiers. Specifically, we define the identifier ``[MSG]'' at the start of a commit message $M=\{m_0,m_1,...,m_i\}$ and ``[FILE]'' for file path $F=\{f_0,f_1,...,f_j\}$ where $i$ and $j$ denote the number of $M$ and $F$ word tokens respectively. Similarly, we also define the identifier ``[CODE]'' at the start of the code to distinguish it from the message and file path. In addition, to distinguish the deleted code statements and the updated statements, we further add the identifiers (``[NEG]''/``[END]'') and (``[POS]''/``[END]'') between the start and end of these deleted/updated statements for distinction. Hence, generally, the code in the commit can be represented as follows: 
\begin{align*}
C=\{c_0...[\mathrm{NEG}]n^{i}_0...[\mathrm{END}][\mathrm{POS}]p^{j}_0...[\mathrm{END}]c_k...\\
[\mathrm{NEG}]n^{i^{'}}_0...[\mathrm{END}][\mathrm{POS}]p^{j^{'}}_0...[\mathrm{END}]...c_l\}
\end{align*}
where $c$ denotes the context statement and $c_0$ is the first token in $c$. Furthermore,
$n^{i}/n^{i^{'}}$ and $p^{j}/p^{j^{'}}$ are the deleted and updated code statements respectively. We utilize segment embedding to embed these segments to ensure the model understands and incorporates information well. Hence, the input embedding representation for the model is constructed by summing the corresponding token, segment and positional embeddings. 

\subsection{Pre-training Tasks}
Apart from text infilling~\cite{lewis2019bart}, we design other five pre-training tasks for commits to pre-train the model and these tasks can be divided into three categories. We use the example from Figure~\ref{fig:commit_example} for illustration. Specifically, the denoising and cross-modal generation tasks are presented in Figure~\ref{fig:four_tasks}, and the contrastive pre-training tasks are presented in Figure~\ref{fig:contrastive_tasks}.
\subsubsection{Denoising Objectives}
Denoising pre-training, which aims at generating the original sequence given the noisy input, is an effective technique for pre-training encoder-decoder models~\cite{lewis2019bart, raffel2020exploring}. We also use it to pre-train \tool. Specifically, we apply a noisy function $N$ to the input sequence $X=\{[\mathrm{CLS}][\mathrm{MSG}]M[\mathrm{FILE}]F[\mathrm{CODE}]C\}$ to get the noisy input defined as $N(X)$, where the noisy function is to corrupt the input sequence by some strategies such as token masking, tokens infilling. The learning process is to ask the model to recover the original sequence from the noisy input and the loss function can be calculated as follows:
\begin{equation}
     \mathcal{L}_{\mathrm{Denoise}}(\theta) = \sum_{t = 1}^{|X|}-\mathrm{log}P_{\theta}(X_t|X_{<t},N(X))
\end{equation}
where $P$ is a generator that generates the $t$-th token given the noisy input $N(X)$, the original sequence before t (i.e., $X_{<t}$) and the model parameters $\theta$. We adopted two strategies to corrupt the semantics of the input sequence (i.e., $X$). 

\textbf{Text Infilling.} It randomly masks spans with a single masked token ``[MASK]''. Then we ask the model to recover the original sequence. Specifically, we set the corruption rate as 15\% in the input sequence and ensure the average span length to 3, followed by a Poisson distribution~\cite{lewis2019bart}. An example is illustrated in Figure~\ref{fig:four_tasks} (a).


\textbf{Graph-guided Token Masking (GTM).} 
Since the input sequence $X$ consists of different components (i.e., commit message, changed code) and they may share some same tokens. Although text infilling helps the model learn better token representations, to further enhance the model in capturing the relations between different components in a commit, we introduce a graph-guided token masking task. A simple example is shown in Figure~\ref{fig:four_tasks} (b). Specifically, by our preliminary observation, we find that there are some critical tokens (e.g., variable names, keywords, file names) that may be shared in commit message $M$ and changed code $C$. Hence, to enhance the semantic mapping, we construct a commit-graph, which links shared tokens between the commit message $M$, file path $F$ and changed code $C$. Then we randomly mask half of the nodes in the graph and require the model to predict these masked nodes by learning the relations from connected neighbors. Through this task, \tool learns to predict the critical tokens in the commit message from the connected nodes in the changed code or vice versa, which improves model learning capacity and alleviates the semantic gap between natural language and code.


\subsubsection{Cross-Modal Generation}
Although denoising objectives ensure that the model learns better token representations, the existing code pre-trained models~\cite{wang2021codet5, guo2022unixcoder} have proved that bimodal generation helps the decoder generate high-quality text output. Hence, we also design two bidirectional conversion generation tasks for GitHub commits to improve the model in generating the natural language text or the code. We formulate both cross-modal generation tasks as follows:

\begin{equation}
    \mathcal{L}_{\mathrm{Gen}}(\theta) = \sum_{t = 1}^{|Y|}-\mathrm{log}P_{\theta}(Y_t|Y_{<t},X_{\mathrm{Gen}})
\end{equation}
where $X_{\mathrm{Gen}}$ is the input sequence and $Y$ is the target sequence for generation.

\textbf{PL2NL Generation.}
As shown in Figure~\ref{fig:four_tasks} (c), this task takes the changed code (i.e. $X_{\mathrm{Gen}} = \{\mathrm{[CLS}][\mathrm{FILE}]F[\mathrm{CODE}]C\}$) as input and requires the model to generate its corresponding commit message $Y=M$ to ensure model produce high-quality natural language texts for downstream natural language generation tasks.


\textbf{PLNL2PL Generation.}
This task aims to generate the updated code snippet based on its commit message and the previous version of the code snippet. Specifically, the previous code snippet before the modification can be expressed as $C^{-} = \{c_0...[\mathrm{NEG}]n^{i^{'}}_0...[\mathrm{END}]...c_l\}$. Hence, the input sequence for this task can be expressed as $X_{\mathrm{Gen}} = \{[\mathrm{CLS}][\mathrm{MSG}]M[\mathrm{FILE}]F[\mathrm{CODE}]C^{-}\}$. The output $Y= \{c_0...[\mathrm{POS}]p^{j^{'}}_0...[\mathrm{END}]...c_l\}$, which is the updated code snippet for the model to generate. We incorporate this task to improve the model in generating better code snippets for downstream code generation tasks.

\subsubsection{Contrastive Learning}
The previous work~\cite{jain2020contrastive} has confirmed that the robustness of code pre-trained models can be enhanced by contrastive learning, we also incorporate it into pre-train \tool.  Generally, it aggregates the similar sequence representation (i.e., $\bm{\widetilde{h}}_i^+$) while pushing away dissimilar representations after encoding the input to an encoder. The loss function is:

\begin{equation}
        \mathcal{L}_{\mathrm{Contra}}(\theta) = \sum_{i=0}^{b-1}-\mathrm{log}\frac{e^{\mathrm{cos}(\bm{\widetilde{h}}_i,\bm{\widetilde{h}}_i^+)/\tau}}{\sum_{j=0}^{b-1}e^{\mathrm{cos}(\bm{\widetilde{h}}_i,\bm{\widetilde{h}}_j)/\tau}}
\end{equation}
where b is batch size, $\tau$ is a temperature hyper-parameter~\cite{wu2018unsupervised} and cos($*$) is the cosine similarity between two vector representations. Specifically, we design two kinds of contrastive strategies for commits.


\textbf{NLPL Alignment.}
As shown in Figure~\ref{fig:contrastive_tasks} (a), NLPL Alignment task takes the vector representations of commit message $M$ with its code $C$ embedded by \tool encoder as a pair of similar sample $(\bm{\widetilde{h}}_i, \bm{\widetilde{h}}_i^+)$ while takes the other representations in the batch as dissimilar samples to further enhance the model capacity of aligning the paired commit message and code.

\textbf{SimCSE.}
 We follow Gao et al.~\cite{gao2021simcse} to put the same input $X=\{[\mathrm{CLS}][\mathrm{MSG}]M[\mathrm{FILE}]F[\mathrm{CODE}]C\}$ with different dropout mask to \tool encoder to get vector representations as the similar samples $(\bm{\widetilde{h}}_i, \bm{\widetilde{h}}_i^+)$ and use other representations in the same batch as dissimilar samples for pre-training. An example is shown in Figure~\ref{fig:contrastive_tasks} (b).


\subsection{Fine-tuning}
We transfer \tool to commit-related tasks at fine-tuning phase. Generally, fine-tuning tasks can be categorized into two classes: understanding tasks and generation tasks. For understanding tasks, we directly feed the source sequence to the encoder and ask the decoder to generate its predicted label. For generation tasks, \tool can be naturally adapted with its encoder-decoder framework to different commit-related generation tasks.

\section{Experimental Setup}


\begin{table}
\centering
  \caption{Results of security patch identification, where * marks the values from Wu et al.~\cite{wu2022enhancing}.}
  \label{tab:patch-identification}
  \scriptsize {\addtolength{\tabcolsep}{0pt}
  \begin{tabular}{l|cccc}
  \hline
    Model& Acc& Pre& Rec & F1\\\hline
    E-SPI & 86.81*&84.49*&91.52*&87.86*\\\hline
    CodeBERT & 88.72& 88.03& 90.85& 89.42\\
    PLBART &92.63&94.77&90.93&92.81\\
    CodeT5-base &93.84&92.54&95.56&94.02\\
    UniXcoder &94.66&94.03&95.88&94.95\\\hline
    Incr-PLBART &94.06&93.07&95.47&94.26\\ \hline
    \tool & \textbf{95.25} &94.93& 96.06 & \textbf{95.49}\\\hdashline
     -w/o SEG  &94.81&94.88&95.22& 95.05  \\\hdashline
    -w/o GTM   &94.56&94.23&95.46&94.84 \\\hdashline
    -w/o PL2NL & 93.84 &93.52 &94.81 &94.16\\
    -w/o PLNL2PL &94.50&94.11&95.46&94.78  \\\hdashline
    -w/o NLPL Align& 94.78& 93.38& \textbf{96.90}& 95.11 \\  
    -w/o SimCSE &95.19&\textbf{96.12}&94.76&95.44 \\
    \hline
  \end{tabular}
  }
  \vspace{-4mm}
\end{table}

\begin{table*}
\centering
  \caption{Smoothed BLEU-4 scores on the commit message generation task. The ``Overall'' column presents the average score over seven programming languages and ``ATOM'' represents experimental results on ATOM dataset.}
  \label{tab:message-generation}
    \scriptsize {\addtolength{\tabcolsep}{0pt}
  \begin{tabular}{l|cccccccc|c}
    \hline
    Model& C & CSharp & Java & JavaScript & PHP & Python & Typescript & Overall & ATOM\\\hline
    Transformer &1.70	&5.24	&10.75	&	9.88&	2.01 &1.34	&	3.70&4.95 &4.76 \\\hline
    CodeBERT &8.75	&	12.39	&	15.14	& 14.45	&	8.80&	13.19&20.55 &	13.32 &10.72\\
    PLBART&11.23	&	15.61	&	15.85	&	15.96	&	11.02	&14.89	&21.60&	15.17 &12.35\\
    CodeT5-base &13.74	&	18.82 &	20.05 &	19.63	 &	12.01 &	18.55 &	23.49 &	18.04 & 13.17\\
    UniXcoder&13.02	 &	18.40	&	20.22	&	20.70&	12.20	&	17.63	&23.63&17.97 & 13.06 \\\hline
    Incr-PLBART &11.98	&16.38	&16.55	&17.33	&	11.24&15.10&21.95 &	15.79 & 13.24\\ \hline
    \tool  &\textbf{15.99}	&	\textbf{21.26} &	22.00	&	26.96	&	\textbf{13.93}	&	\textbf{19.50}	&	\textbf{24.56} &	\textbf{20.60} & \textbf{17.85}\\\hdashline
     -w/o SEG  &15.65	&	21.12	&	22.25	&	26.10	&	13.38	&	19.08	&	23.17 &	20.11 & 16.94\\\hdashline
    -w/o GTM  &15.85	&	20.46	&	\textbf{22.91}	&	26.31	&	13.12	&	18.65	&	23.04 &	20.05  & 16.27\\\hdashline
    -w/o PL2NL &14.54	&	18.66	&	20.33	&	22.89	&	12.21	&	18.00	&	22.08 &	18.39 & 16.25\\
    -w/o PLNL2PL &15.86	&	21.51	&	\textbf{22.91}	&	\textbf{26.97}	&	13.29	&	19.40	&	22.20 &	20.31 &  16.13 \\\hdashline
    -w/o NLPL Align &15.94	&	20.62	&	22.53	&	26.63	&	12.87	&	19.15	&	22.98 &	20.10 & 17.33\\
    -w/o SimCSE &15.95	&	20.67	&	22.54	&	26.85	&	13.62	&	19.11	&	23.06	&20.26& 17.28\\
    \hline
  \end{tabular}
  }
\vspace{-4mm}
\end{table*}

In this section, we first introduce the evaluation tasks and then introduce the compared baselines. The details of pre-training and fine-tuning settings can be found in Appendix~\ref{sec:pre-training-setting} and Appendix~\ref{sec:fine-tune-setting}.
\subsection{Evaluation Tasks}
We select one understanding task (i.e., security patch identification) and three generation tasks (i.e., commit message generation, positive code statements generation, and updated code snippet generation) as  fine-tuning tasks. 

\noindent \textbf{Security Patch Identification.} This task aims to identify whether a commit fixes a software vulnerability or not. It has been extensively researched with some neural models (e.g., SPI~\cite{zhou2021spi}, E-SPI~\cite{wu2022enhancing}, PatchRNN~\cite{wang2021patchrnn}). We use the same dataset provided by E-SPI~\cite{wu2022enhancing} with the same train-validation-test split for evaluation. 

\noindent  \textbf{Commit Message Generation.} This task targets generating a commit message to summarize the changed code in natural language. Apart from evaluating the widely used open-source dataset ATOM~\cite{liu2020atom} for Java programming language, we also used the constructed fine-tuning dataset from our benchmark (See Table~\ref{tab:data-statistics}) across \textbf{7} languages for evaluation. We split the data into train/validation/test based on ``project'' to evaluate. Followed by Tao et al.~\cite{tao2021evaluation}, we use smoothed BLEU-4 as the evaluation metric. 

\noindent \textbf{Positive Code Statements Generation.} We propose this new commit-related task, which targets generating positive statements marked by ``+'', that takes the commit message, and file path with the code snippet before modification as input. We only consider the commits that have consecutive statement modifications for evaluation. Specifically, if a commit in our constructed fine-tuning dataset from Table~\ref{tab:data-statistics} has multiple non-consecutive modifications (a simple example is shown in Appendix Figure~\ref{fig:code_gen_source_2}, which removes two non-consecutive lines of statements), we remove these samples. This task is similar to source code edit~\cite{chakraborty2021multi}. Compared with it, which generated patched code statements based on buggy code statements, ours aim to generate updated code statements from the previous version of code statements. 



\noindent \textbf{Updated Code Snippet Generation.} We further propose a more challenging task that requires the model to generate the completed updated code snippet. Different from positive code statements generation, this task needs to locate the position of the updated statements first and then generate the completed code snippet. We include these commits that have multiple inconsecutive modifications (see an example in Appendix Figure~\ref{fig:code_gen_source_2}). This task is valuable for software developers to generate the code that meets the requirement based on the previous version of code as well as the demand in natural language. We use our fine-tuning dataset in Table~\ref{tab:data-statistics} for evaluation.




The statistics of the used datasets for these fine-tuning tasks are presented in Appendix~\ref{sec:fine-tune-setting}. 

\vspace{-2mm}
\subsection{Baselines}
For the understanding task of security patch identification, since we use the dataset from E-SPI~\cite{wu2022enhancing} with the same train-validation-test split, we directly report the best results from their paper as one of our baselines. For generation tasks, we add one supervised baseline Transformer model~\cite{DBLP:conf/nips/VaswaniSPUJGKP17}, which is trained on fine-tuning commit dataset without pre-training to verify the effectiveness of pre-training techniques. Furthermore, we compare \tool with four state-of-the-art pre-trained models for code (i.e., CodeBERT~\cite{feng2020codebert}, PLBART~\cite{ahmad2021unified}, CodeT5-base~\cite{wang2021codet5} and UniXcoder~\cite{guo2022unixcoder}), which are trained on a large corpus of code functions to validate the effectiveness of utilizing commit data to pre-train a model for commit-related tasks. We directly employ these released pre-trained models with default configurations for comparison. 
In addition, since our model adopts the parameters of PLBART, which is trained by text infilling, to validate the effectiveness of other designed pre-training tasks in \tool, we further add one baseline Incr-PLBART and it is incrementally trained from PLBART on our collected commit benchmark using text infilling.

\begin{table*}
\centering
  \caption{BLEU-4 scores and exact match (EM) accuracies for positive code statements generation task.}
  \label{tab:single-line}
    \scriptsize {\addtolength{\tabcolsep}{-4pt}
  \begin{tabular}{l|cccccccccccccccc}
    \hline
    \multirow{2}{*}{Model}& \multicolumn{2}{c}{C} & \multicolumn{2}{c}{CSharp} & \multicolumn{2}{c}{Java} & \multicolumn{2}{c}{JavaScript} & \multicolumn{2}{c}{PHP} & \multicolumn{2}{c}{Python} & \multicolumn{2}{c}{TypeScript} & \multicolumn{2}{c}{Overall}  \\ \cline{2-17}
         & EM  & BLEU-4      & EM & BLEU-4      & EM & BLEU-4      & EM & BLEU-4      & EM & BLEU-4      & EM & BLEU-4      & EM & BLEU-4      & EM & BLEU-4\\ \hline
    Transformer &0.02 &	6.10&	0.00 &	8.32 &	0.00 &	13.26 &	0.00 &	9.59 &	0.00 &	5.14 &	0.00 &	6.95 &	0.02 &	15.37 &	0.01 &	9.25\\\hline
    CodeBERT &6.25 &	39.94 &	20.24 &	47.08 &	30.33 &	48.04 &	18.54 &	50.85 &	6.63 &	48.57 &	15.86 &	44.74 &	38.79 &	53.30 &	19.52 &	47.50\\
    PLBART&7.92	&47.56 &	19.36 &	50.75 &	31.07 &	52.19 &	18.86 &	51.39 &	8.23 &	50.81&	16.26&	48.83&	37.40&	53.73&	19.87 &	50.75\\
    CodeT5-base &13.41 &	50.94 &	25.50 &	52.98 &	38.80 &	56.14 &	23.71 &	59.10 &	10.73 &	49.07 &	20.61 &	51.02 &	43.84 &	60.51 &	25.23 &	54.25 \\
    UniXcoder&14.60&	53.07&	26.26&	56.20&	35.80&	57.11&	23.48	&60.62&	13.07&	53.93&	22.71&	52.51&	42.86&	62.07&	25.54&	56.50 \\\hline
    Incr-PLBART &9.93 &	48.12 &	21.46 &	51.21 &	33.29 &	52.80 &	21.42 &	54.44 &	9.79 &	51.00&	18.41 &	48.73 &	38.88 &	56.38 &	21.88 &	51.81\\ \hline
    \tool  &\textbf{16.64} &	\textbf{55.69} &	\textbf{28.26} &	\textbf{58.04} &	\textbf{41.02} &	\textbf{61.05} &	\textbf{34.12} &	60.50 &	\textbf{15.21} &	56.80 &	\textbf{25.61} &	56.17 &	\textbf{47.16} &	61.97 &	\textbf{29.72} &	\textbf{58.60}\\\hdashline
     -w/o SEG  &15.50 &	55.13 &	27.38 &	56.81 &	40.60 &	59.99 &	31.79 &	\textbf{62.25} &	13.79 &	55.35 &	24.02 &	56.10 &	46.63 &	\textbf{62.28} &	28.53 &	58.27  \\\hdashline
    -w/o GTM  &15.78 &	53.83 &	26.72 &	56.39 &	40.14 &	60.18 &	31.15 &	60.97 &	14.51 &	55.16 &	24.46 &	55.88 &	46.80 &	60.37 &	28.51 &	57.54  \\\hdashline
    -w/o PL2NL &15.36 &	55.21 &	26.66 &	56.94 &	39.90 &	59.71 &	27.19 &	58.28 &	13.40 &	\textbf{57.37} &	23.60 &	54.69 &	44.21 &	61.67 &	27.19 &	57.70 \\
    -w/o PLNL2PL &14.22& 53.93 &	24.86 &	57.14 &	36.12 &	59.28 &	24.15 &	60.64 &	12.83 &	56.01 &	23.26 &	54.49 &	43.74 &	61.84 &	25.60 &	57.62  \\\hdashline
    -w/o NLPL Align &15.30 &	54.99 &	27.24 &	56.44 &	40.18 &	59.99 &	30.35 &	59.98 &	14.08 &	56.93 &	24.17 &	\textbf{56.20} &	45.24 &	62.00 &	28.08 &	58.08\\
    -w/o SimCSE &16.07 &	53.44 &	28.04 &	57.57 &	40.18 &	59.69 &	32.52 &	61.67 &	14.86 &	56.23 &	24.43 &	55.79 &	46.38 &	59.86 &	28.93 &	57.75\\
    \hline
  \end{tabular}
  }
\vspace{-1mm}
\end{table*}
\section{Experimental Results and Analysis}

In this section, we present the experimental results with the analysis.

\subsection{Compared with Baselines}
We compare the results of these evaluation tasks with the baselines in Table~\ref{tab:patch-identification}, Table~\ref{tab:message-generation}, Table~\ref{tab:single-line} and Table~\ref{tab:code-snippet} respectively. We can conclude the following findings: 1) Compared with the supervised techniques such as Transformer, the pre-trained models improve the performance significantly over these tasks. 2) Compared with the pre-trained models in code, \tool outperforms them, especially on the generation tasks significantly. We attribute the improvements to the used large amount of commit data. Specifically, we utilize over \textbf{7} million commits to pre-train a commit-related model for downstream commit-related tasks and this model is domain-specific. 3) Compared with Incr-PLBART, our improvements are also significant, which confirms that except for text infilling, the other pre-training tasks in \tool are also beneficial for the improvements.

\begin{table*}
\centering
\scriptsize {\addtolength{\tabcolsep}{-4pt}
\centering
  \caption{BLEU-4 scores and exact match (EM) accuracies for updated code snippet generation task.}
  \label{tab:code-snippet}
    \scriptsize {\addtolength{\tabcolsep}{0pt}
  \begin{tabular}{l|cccccccccccccccc}
    \hline
    \multirow{2}{*}{Model}& \multicolumn{2}{c}{C} & \multicolumn{2}{c}{CSharp} & \multicolumn{2}{c}{Java} & \multicolumn{2}{c}{JavaScript} & \multicolumn{2}{c}{PHP} & \multicolumn{2}{c}{Python} & \multicolumn{2}{c}{TypeScript} & \multicolumn{2}{c}{Overall}  \\ \cline{2-17}
    &  EM  & BLEU-4      & EM & BLEU-4      & EM & BLEU-4      & EM & BLEU-4      & EM & BLEU-4      & EM & BLEU-4      & EM & BLEU-4      & EM & BLEU-4\\ \hline
    Transformer &0.00 &	8.36 &	0.00	 &17.27 &	0.00 &	17.83 &	0.00 &	9.71 &	0.00 &	10.20 &	0.00 &	7.38 &	0.01 &	23.80 &	0.00 &	13.51\\\hline
    CodeBERT &6.16 &	48.63 &	13.11 &	48.83 &	18.10 &	47.64 &	14.12 &	45.59 &	8.71 &	51.59 &	10.39 &	43.14 &	24.89 &	56.71 &	13.64 &	48.88\\
    PLBART&8.75 &	51.99 &	14.77 &	56.56 &	22.28 &	51.85 &	15.98 &	50.28 &	10.67 &	55.47 &	13.04 &	49.39 &	27.73 &	62.57 &	16.17 &	54.02\\
    CodeT5-base &8.96 &	53.42 &	12.96 &	56.48 &	22.23 &	52.73 &	17.26 &	48.92 &	7.72 &	53.11 &	10.72 &	46.81 &	24.02 &	58.15 &	14.84 &52.80 \\
    UniXcoder&11.25 &	52.90 &	15.04 &	53.38 &	24.09 &	54.05 &	20.15 &	54.88 &	13.86 &	55.64 &	18.12 &	53.03 &	30.97 &	61.08 &	19.07&	54.99 \\\hline
    Incr-PLBART &10.52 &	55.08 &	16.47 &	56.79 &	24.32 &	55.68 &	16.53 &	54.21 &	12.05 &	57.51 &	14.72 &	50.87 &	29.86 &	64.39 &	17.78 &	56.36\\ \hline
    \tool  &\textbf{17.18} &	56.80 &	\textbf{22.45} &	\textbf{57.27} &	\textbf{31.30} &	53.99 &	\textbf{29.14} &	\textbf{55.52} &	17.37 &	59.16 &	19.96 &	\textbf{54.00} &	\textbf{35.43} &	\textbf{66.64} &	\textbf{24.69} &	\textbf{57.63}\\\hdashline
     -w/o SEG  &16.54 &	56.05 &	21.67 &	55.75 &	30.36 &	\textbf{55.40} &	28.36 &	55.49 &	17.13 &	\textbf{59.69} &	\textbf{19.97} &	51.51 &	35.24 &	64.86 &	24.18 &	56.96 \\\hdashline 
    -w/o GTM  &16.19 &	56.54 &	21.49 &	54.77 &	30.20 &	54.73 &	27.48 &	54.42 &	16.53 &	57.57 &	18.94 &	51.64 &	34.30 &	63.00 &	23.59&	56.10\\\hdashline
    -w/o PL2NL &16.60 &	56.61 &	21.56 &	56.49 &	30.51 &	54.74 &	27.82 &	55.12 &	16.92 &	57.86 &	19.53 &	50.89 &	34.53 &	64.86 &	23.92 &	56.65 \\
    -w/o PLNL2PL &12.35 &	52.72 &	17.60 &	53.26 &	24.99 &	49.99 &	21.61 &	50.41 &	14.65 &	58.02 &	18.13 &	51.23 &	31.80 &	60.28 &	20.16&	53.70 \\\hdashline
    -w/o NLPL Align &16.81 &	56.73 &	21.80 &	57.17 &	31.29 &	54.67 &	28.48 &	54.28 &	\textbf{17.45} &	59.84 &	19.94 &	53.50 &	35.12 &	62.42 &	24.41 &	56.94\\
    -w/o SimCSE &17.14 &	\textbf{57.29} &	21.95 &	55.71 &	30.50 &	55.10 &	28.70 &	54.74 &	17.22 &	59.52 &	19.74 &	51.80 &	35.18 &	65.77 &	24.35 &	57.13\\
    \hline
  \end{tabular}
  }
  }
    \vspace{-4mm}
\end{table*}

\vspace{-2mm}
\subsection{Model Analysis}
We further conduct an ablation study to investigate each pre-training task to the final performance across these fine-tuning tasks and the results are reported in the last line of Table~\ref{tab:patch-identification}, Table~\ref{tab:message-generation}, Table~\ref{tab:single-line} and Table~\ref{tab:code-snippet} respectively. Through analysing the results, we have the following findings: 1) Segment embedding is useful in \tool, although it is not widely used in code pre-trained models. We believe it is due to commits usually having more complex components (e.g., commit message, deleted/updated statements), hence segment embedding helps the model incorporate each component well. 2) Two cross-modal generation tasks (i.e., PL2NL and PLNL2PL) provide significant improvements to the generation tasks. Taking the updated code snippet generation task in Table~\ref{tab:code-snippet} as an example, after removing PLNL2PL, the overall BLEU-4 score drops from 57.63 to 53.70 across \textbf{7} programming languages. 3) Apart from the cross-modal generation tasks, graph-guided token masking (GTM) is more critical than the remaining tasks on three generation tasks. It demonstrates that by constructing a commit-graph to predict the masked nodes by their connected nodes in the graph, the semantic gap between the message (NL) and changed code (PL) is effectively alleviated. 4) Both contrastive learning pre-training tasks are effective in improving performance. Overall, we conclude that each pre-training task improves \tool.

\vspace{-2mm}

\subsection{Case Study}
\begin{figure}[t]
     \centering
     \includegraphics[width=\columnwidth]{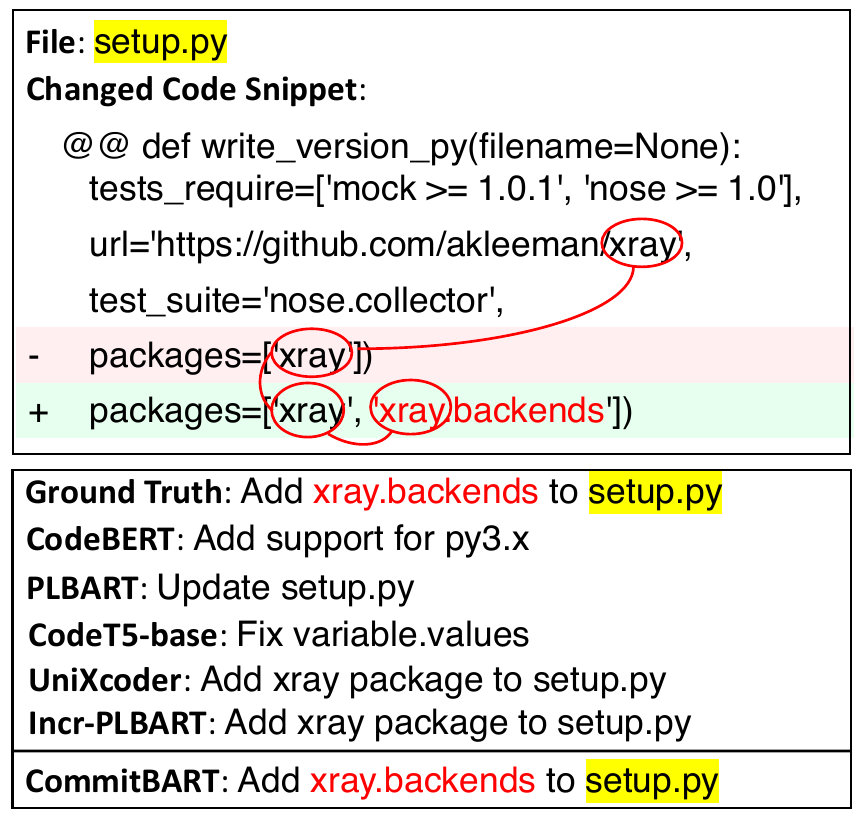}
    \caption{One example from Python for the task of commit message generation where the commit id is
    \href{https://github.com/thadncs/https-github.com-pydata-xarray/commit/a68e10e866f24b014dc0ef8a670a552cf5966b15}{966b15}.}
     \label{fig:case_study_one}
     \vspace{-6mm}
\end{figure}

We also conduct a case study including an example from commit message generation with the generated results from different models to intuitively demonstrate the effectiveness of \tool. It is shown in Figure~\ref{fig:case_study_one} and more examples of different models for these commit-related generation tasks can be found in Appendix~\ref{sec:more-case-study}. From the changed code in this example, we find that this commit is to add a package namely ``xray.backends''. Its commit message (i.e., ground-truth) ``Add xray.backends to setup.py'' further confirms it. The results produced by UniXcoder and Incr-PLBART are better than other baselines. However, both models fail to generate the accurate package name (i.e., ``xray.backends''). In contrast, \tool generates the accurate package name and the generated output is the same as the ground-truth. We attribute to the designed graph-guided token masking pre-training task, which helps the model capture the critical semantic information in a commit.

\section{Conclusion}
\vspace{-1mm}
Due to the lack of large-scale commit data, we first collect a benchmark (over \textbf{7.99} million commits) across \textbf{7} programming languages and make it public for follow-up researchers. Based on this benchmark, we present \tool, a pre-trained encoder-decoder model for GitHub commits. The model is pre-trained via three categories (i.e., denoising objectives, cross-modal generation and contrastive learning) for six pre-training tasks to learn commit representations. We are the first to present a large pre-trained model for GitHub commits and further, unify a ``commit intelligence'' framework with one commit-related understanding task and three generation tasks. Extensive experiments confirm \tool significantly outperforms previous works on these tasks. Further ablation study also reveals the effectiveness of each pre-training task. We encourage more commit-related tasks to merge into our framework.
\section{Limitations}
While \tool is able to achieve superior performance for different code-related tasks. We claim that it still has the following limitations:

\noindent First, for the downstream task of updated code snippet generation, we just use BLEU-4 and exact match as the evaluation metrics. However, recent studies~\cite{chen2021evaluating} have proved the deficiencies in match-based metrics such as BLEU for code and further proposed the execution-based metric (i.e., pass@k) to evaluate the functional correctness of the generated code. Although pass@k can be used for the testset (i.e., HumanEval~\cite{chen2021evaluating}), which evaluates the correctness of the generated code based on the test cases, this operation of designing test cases for the task of updated code snippet are not applicable because these generated code snippets from the previous version of code are complex and they cannot compile without the well-configured operating environment.  

\noindent Second, we just select the first sentence as the target for commit message. It will lose some useful information in some cases where the left sentences are also important to this changed code. We extract the first sentence so that we can accelerate the training and inference process.

\bibliography{anthology,main}
\bibliographystyle{acl_natbib}

\appendix
\begin{table*}[ht!]
\centering
\small
  \caption{The statistics of the fine-tuning dataset for commit message generation, updated code snippet generation and positive code statements generation. The first row is the dataset used for the first two tasks while second row is the dataset for the last task.}
  \label{tab:fine-data-statistics}
  \begin{tabular}{crrrrrrrr}
    \hline
    Fine-tune&C & CSharp & Java & JavaScript & PHP & Python & Typescript & Total\\
    \hline
    Train & 51,504& 45,425 & 63,189 & 68,338 &78,450  &70,137 & 51,526  &428,569 \\
    Validation &8,902 & 5,501 & 7,512  &8,100  &9,239  & 8,326& 6,172 & 53,752\\
    Test &  11,518& 10,976 &  10,425&  10,626& 11,541 & 11,039& 10,064 & 76,189\\ \hline
    Positive-Train & 20,746 &  19,726& 27,619 & 31,809 &35,015 & 28,198 & 26,325& 189,438\\
    Positive-Validation & 2,320 &  2,561& 2,580 & 2,655 & 2,490 & 2,363& 2,957 & 17,926\\
    Positive-Test & 2,321 & 2,562 & 2,580  & 2,656 &  2,491& 2,364&  2,958& 17,932 \\
    \hline
  \end{tabular}
\end{table*}

\begin{figure}[tp]
     \centering
     \includegraphics[width=\columnwidth]{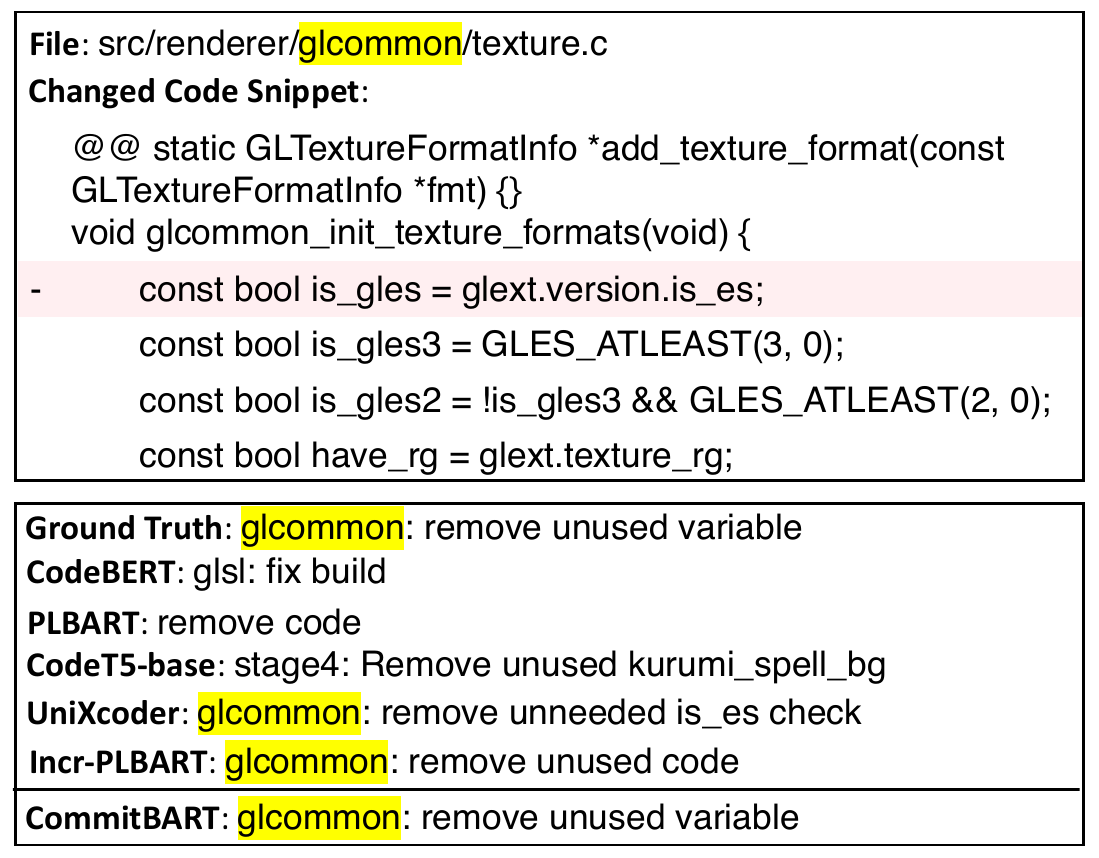}
    \caption{One example from C for the task of commit message generation where the commit id is
    \href{https://github.com/Kinoli7/taisei/commit/6072a720465d60ca84d1a9825b132a1f138457d5}{8457d5}.}
     \label{fig:case_study_two}
\end{figure}

\section{Pre-training Settings}\label{sec:pre-training-setting}
We follow the architecture of BART~\cite{lewis2019bart}, which is made up of a 6-layer Transformer encoder and a 6-layer decoder with the dimension of 768 and 12 heads ($\sim$140M parameters). We adopt the parameters of PLBART~\cite{ahmad2021unified} to initialize the pre-trained model and add an additional segment embedding layer to encode different segment identifiers. We also adopt the vocabulary set of PLBART, which contains 50K subtokens and further add 5 different segment identifiers into the vocabulary set to represent different components in a commit. We set the maximum input sequence length to 512 and use the mixed precision of FP16 to accelerate the pre-training process. We set the batch size to 512 and employ the AdamW optimizer to update the model parameters with a learning rate of 2e-4. We pre-train the model on one DGX server, which has 8 NVIDIA Tesla V100 with 32GB memory. The total steps are set to 80K where the categories of denoising objectives, cross-modal generation and contrastive learning take up 60\%, 30\%, and 10\% respectively. One category has two different pre-training tasks and each of them accounts for half of the steps. The total time for the pre-training process is about 60 hours. We follow Guo et al.~\cite{guo2020graphcodebert} to sample each batch from the same programming
language according to a distribution $\{q_i\}_{i=1...N}$, where $n_i$ is the number of examples for $i$-th programming language and $\alpha=0.7$ to alleviate the bias towards high-resource languages.

\begin{equation}
q_i=\frac{p^\alpha_i}{\sum_{j=1}^{N}p^\alpha_j} \ , \ p_i=\frac{n_i}{\sum_{k=1}^{N}n_k}
\end{equation}

\begin{figure*}[t!]
     \centering
     \begin{subfigure}{0.49\textwidth}
         \centering
         \includegraphics[width=\textwidth]{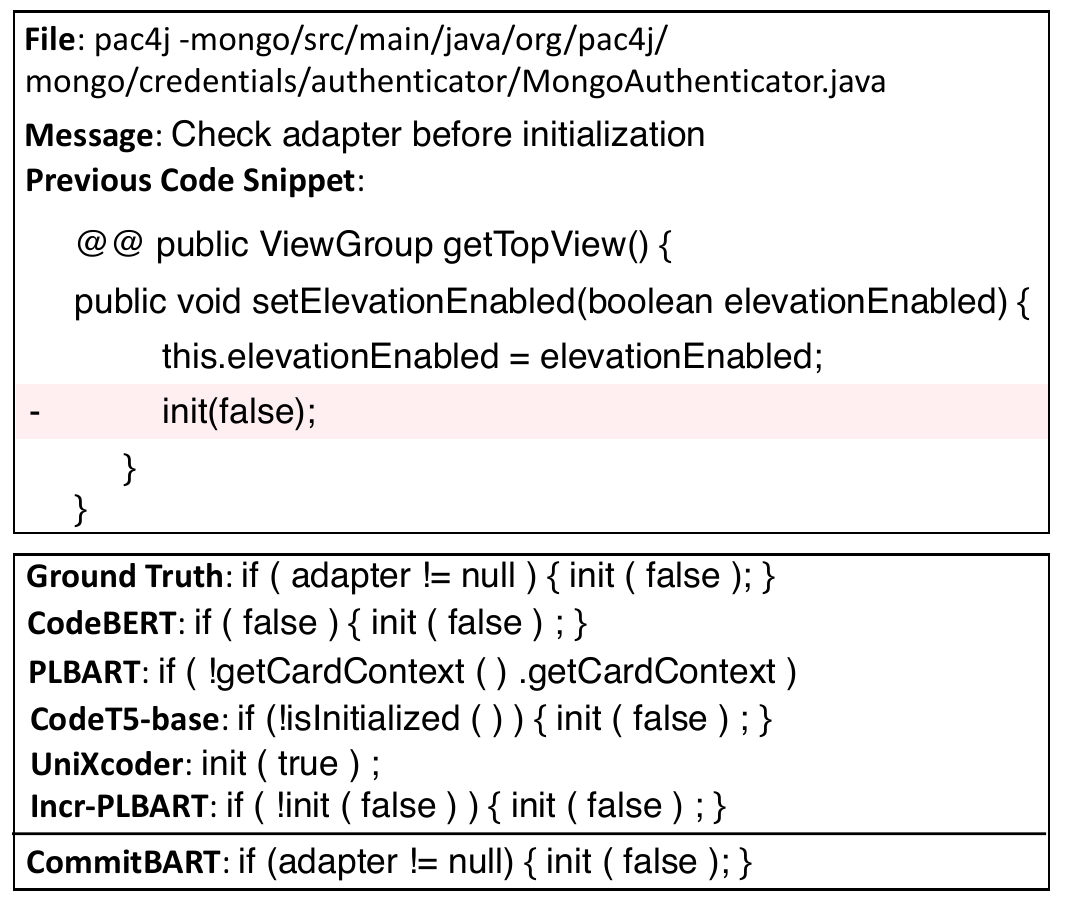}
         \caption{The commit id is \href{https://github.com/abhirocks1211/CardStackView/commit/86094473ab4a3562587235f243cf95c0f1054122}{054122}.}
         \label{fig:single_example_1}
     \end{subfigure}
     \hfill
     \begin{subfigure}{0.49\textwidth}
         \centering
         \includegraphics[width=\textwidth]{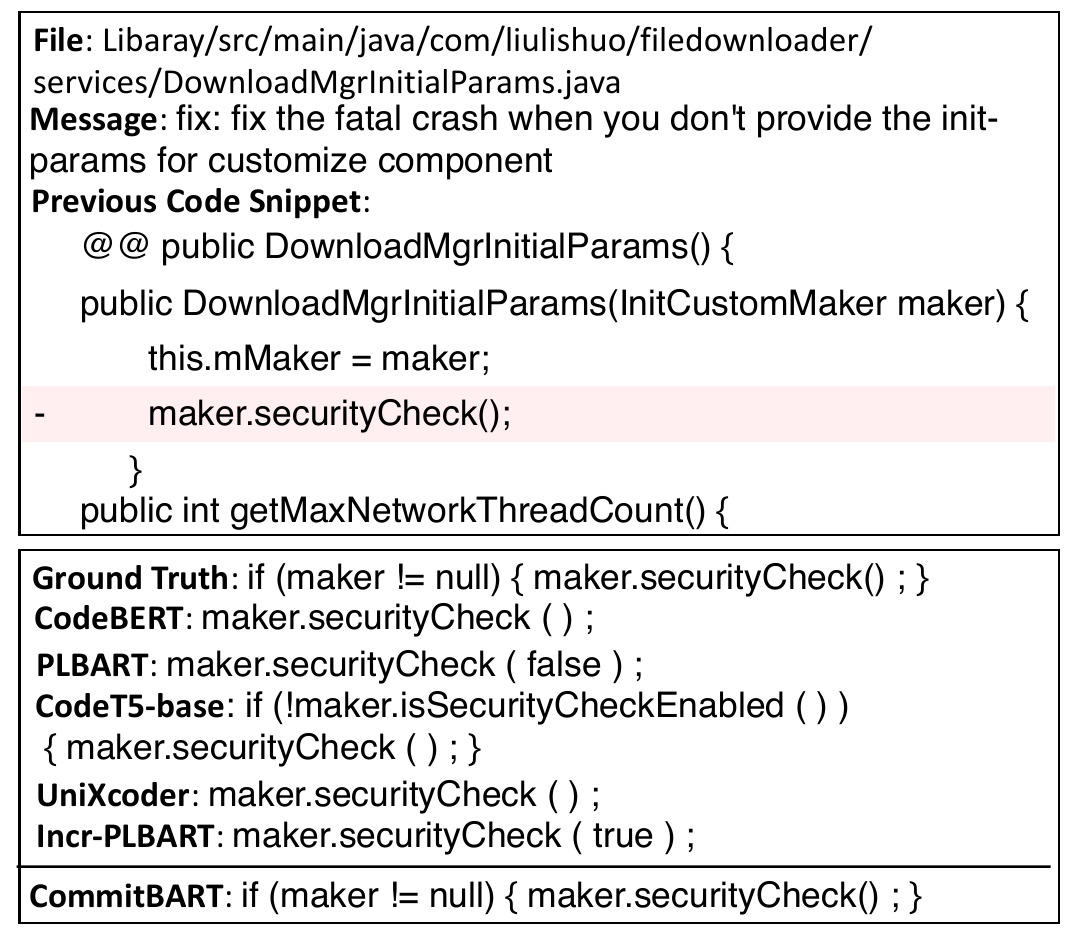}
         \caption{The commit id is \href{https://github.com/renxuetao/FileDownloader/commit/447ea4dc7c19491b2b8bb2bd92418cbcad99cfb9}{99cfb9}.}
         \label{fig:single_example_2}
     \end{subfigure}
        \caption{Two examples from Java for the task of positive code statements generation.}
        \label{fig:single_examples}
\end{figure*}

\section{Fine-tuning Settings}\label{sec:fine-tune-setting}
\subsection{Understanding Task}
For the understanding task of security patch identification, we directly utilize the released dataset from Wu et al.~\cite{wu2022enhancing}, which consists of 26500 training samples, 3301 validation samples and 3294 test samples for evaluation. We employ AdamW optimizer to fine-tune \tool with a 5e-5 learning rate and the batch size 32 for 4K steps. We set the maximum input sequence length to 512 and the target sequence length to 5, which includes the start token ``[CLS]'', the task prefix ``security patch'', the label ``True'' or ``False'' and the end token ``[EOS]''. 

\subsection{Generation Tasks}
For the generation task of commit message generation, apart from the dataset used in ATOM~\cite{liu2020atom}, we also utilize the fine-tuning dataset from our benchmark (See Table~\ref{tab:data-statistics}). Specifically, we split it into train/validation/testset based on the ``project'' with a ratio of 75\%:10\%:15\% for evaluation. For the task of updated code snippet generation, we also use this divided data for evaluation.
For the task of positive code statements generation, we extract the consecutive modification samples from the constructed fine-tuning dataset for evaluation. The statistics of the train, validation and test are shown in Table~\ref{tab:fine-data-statistics}. For each task, we utilize AdamW optimizer to fine-tune \tool with a 5e-5 learning rate of batch size 32 for 10K steps. Furthermore, we set the early stop based on the validation loss. The maximum input sequence length is set to 512 for these tasks, while the target sequence is set to 150 for commit message generation, 512 for updated code snippet generation and 300 for positive code statements generation respectively.

\section{Case Study}\label{sec:more-case-study}
In this section, we provide more examples to demonstrate the effectiveness of \tool. 
\subsection{Commit Message Generation}
We further provide one example of commit message generation. From Figure~\ref{fig:case_study_two}, we can see that its commit message is to delete an unused variable and the changed code is to delete the statement (i.e., ``const bool is\_gles = glext.version.is\_es''). By comparing the results produced by different models, we find that \tool can generate a better result.

\subsection{Positive Code Statements Generation}
We provide two examples for the task of positive code statements generation in Figure~\ref{fig:single_example_1} and Figure~\ref{fig:single_example_2} respectively. The input consists of the file path, the commit message and the previous code snippet. We ask the model to generate the positive code statements marked as ``+'' in a commit. We can see that both commits aim to add a security check to avoid code crashes. Compared with the results produced by different models, \tool generates accurate positive code statements.

\subsection{Updated Code Snippet Generation}
\begin{figure*}
     \centering
         \centering
     \begin{subfigure}{0.49\textwidth}
         \centering
         \includegraphics[width=\textwidth]{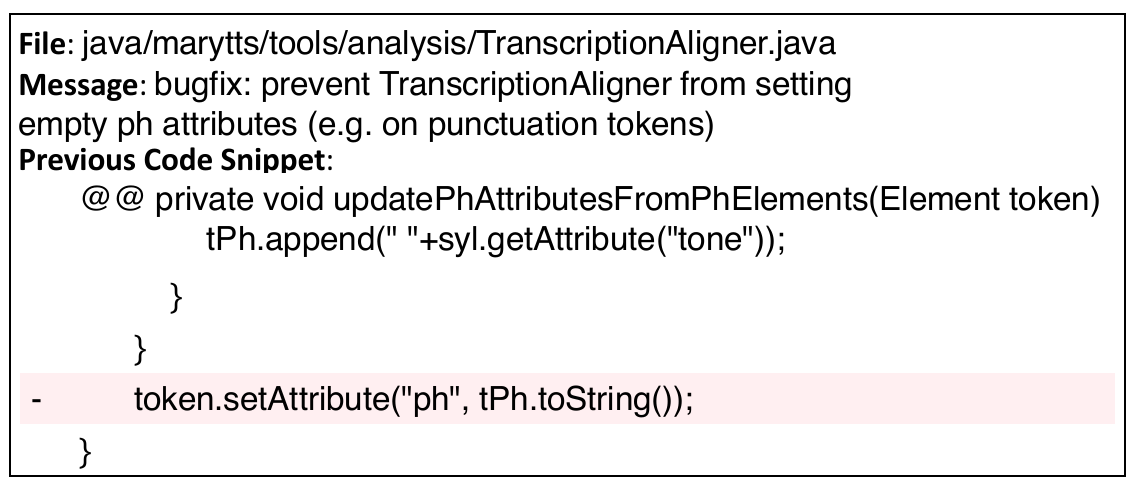}
         \caption{The input for models to generate the updated code snippet.}
         \label{fig:code_gen_source}
     \end{subfigure}
     \hfill
     \begin{subfigure}{0.49\textwidth}
         \centering
         \includegraphics[width=\textwidth]{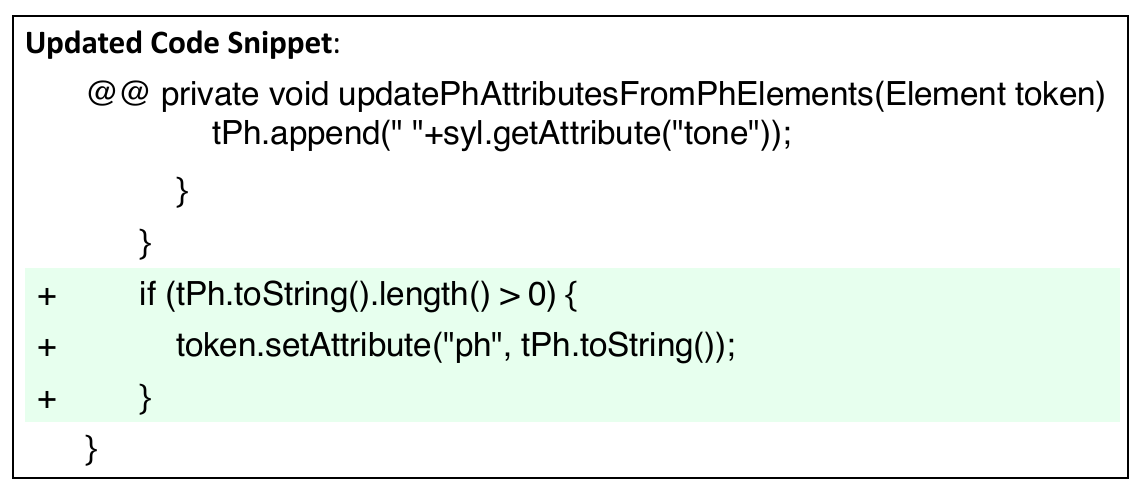}
         \caption{The updated code snippet (i.e., ground-truth).}
         \label{fig:code_gen_ground}
     \end{subfigure}
     \begin{subfigure}{0.49\textwidth}
         \centering
         \includegraphics[width=\textwidth]{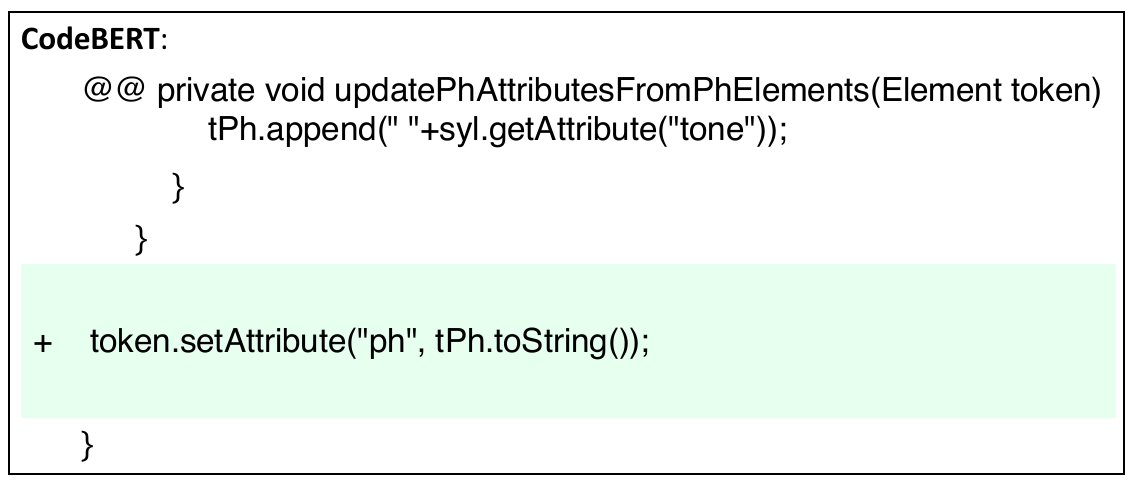}
         \caption{The result provided by CodeBERT.}
         \label{fig:code_gen_codebert}
     \end{subfigure}
    \hfill
     \begin{subfigure}{0.49\textwidth}
         \centering
         \includegraphics[width=\textwidth]{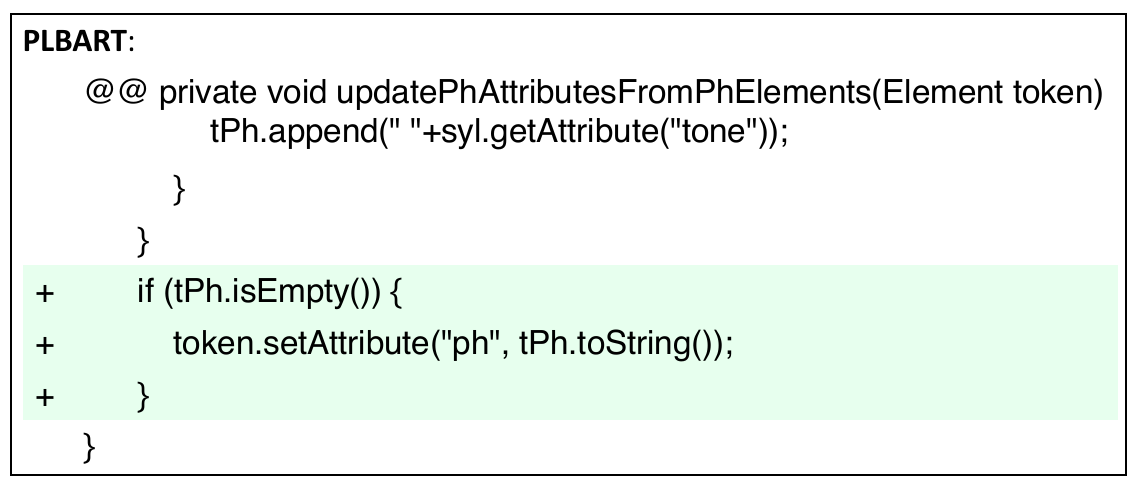}
         \caption{The result provided by PLBART.}
         \label{fig:code_gen_plbart}
     \end{subfigure}
        \hfill
     \begin{subfigure}{0.49\textwidth}
         \centering
         \includegraphics[width=\textwidth]{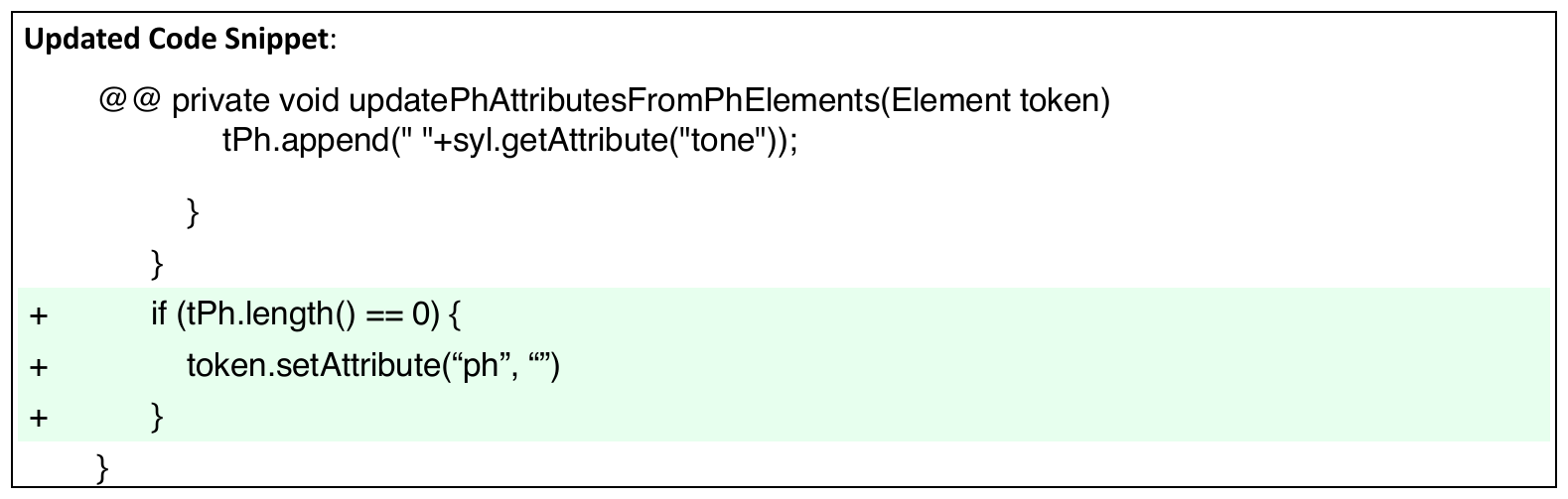}
         \caption{The result provided by CodeT5-base.}
         \label{fig:code_gen_codet5}
     \end{subfigure}
         \hfill
        \begin{subfigure}{0.49\textwidth}
         \centering
         \includegraphics[width=\textwidth]{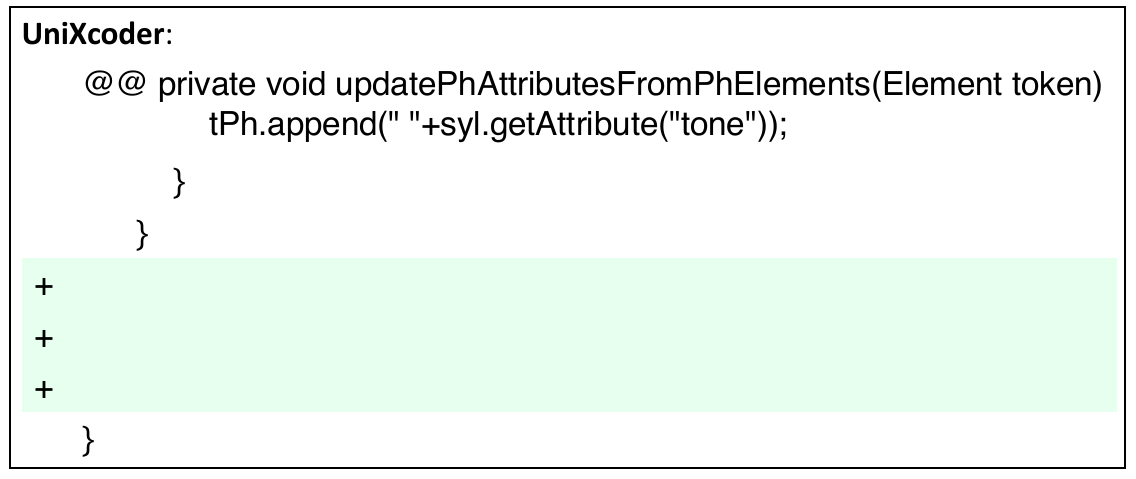}
         \caption{The result provided by UniXcoder.}
         \label{fig:code_gen_unixcoder}
     \end{subfigure}
         \hfill
        \begin{subfigure}{0.49\textwidth}
         \centering
         \includegraphics[width=\textwidth]{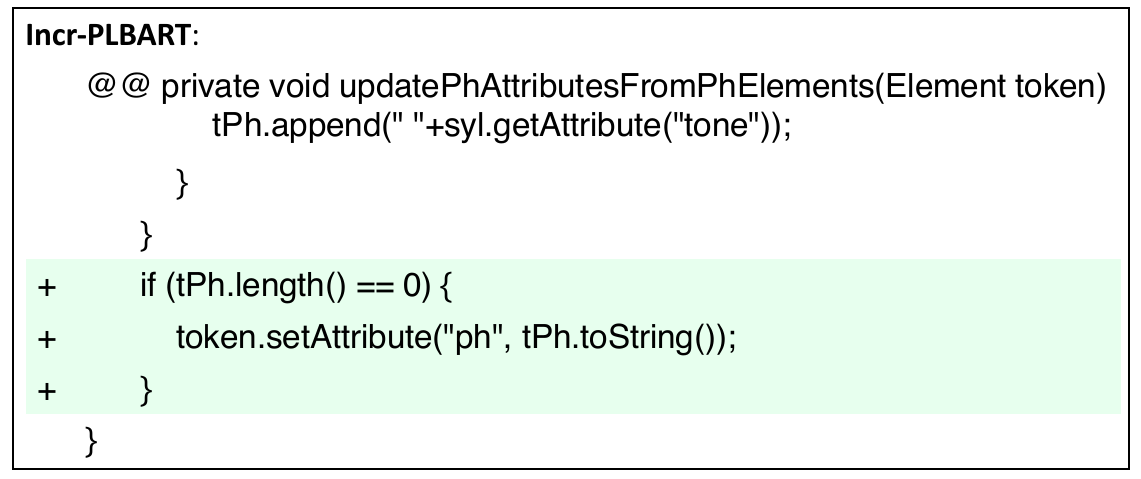}
         \caption{The result provided by Incr-PLBART.}
         \label{fig:incr_plbart}
     \end{subfigure}
         \hfill
     \begin{subfigure}{0.49\textwidth}
         \centering
         \includegraphics[width=\textwidth]{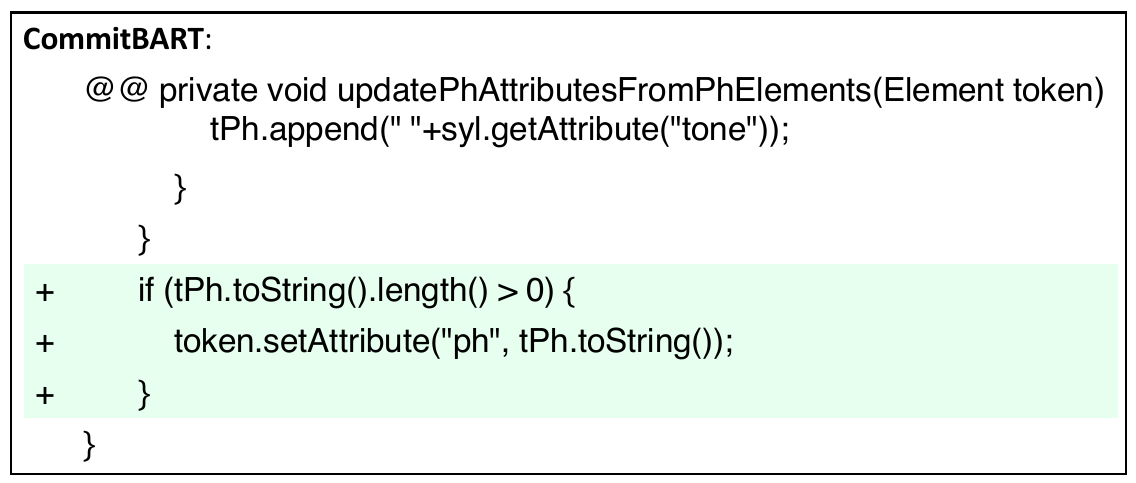}
         \caption{The result provided by CommitBART.}
         \label{fig:code_gen_commitbart}
        \end{subfigure}
         \caption{One example from Java for the task of updated code snippet generation where the commit id is \href{https://github.com/marytts/marytts/commit/0f5b3efb61a6762fed204b776f7e1730c20aa027}{0aa027}.}
        \label{fig:code_gen}
\end{figure*}

We provide two examples for the task of updated code snippet generation. Similar to the task of positive code statements generation, the input includes the file path, the commit message and the previous code snippet. For the first example, the input is presented in Figure~\ref{fig:code_gen_source} and the ground-truth for this example is presented in Figure~\ref{fig:code_gen_ground}. By Figure~\ref{fig:code_gen_source}, we can get that the updated code snippet needs to add a variable safety check to avoid ``ph'' as empty. The results produced by different models are presented in Figure~\ref{fig:code_gen_codebert} to Figure~\ref{fig:code_gen_commitbart} accordingly. We can observe that PLBART and Incr-PLBART produce better results than other baseline models for this example, however, they both misunderstand the semantics of the commit message and produce the exact opposite results (i.e., ``if (tPh.isEmpty())'' and ``if (tPh.length() == 0)''). In contrast, \tool can capture the semantics well and produce the same code snippet with the ground-truth. For another example, which is shown in Figure~\ref{fig:code_gen_2}, we can see that this commit removes the variable ``isContinuousIntegration'' and further deletes its usage in the following statement. The modification is non-consecutive. The results produced by CodeBERT and CodeT5 only copy the original statements while the results produced by the remaining baselines remove the variable declaration, but fail to delete its usage. In contrast, \tool can produce the same code snippet with the ground-truth.

\begin{figure*}
     \centering
     \begin{subfigure}{0.8\textwidth}
         \centering
         \includegraphics[width=\textwidth]{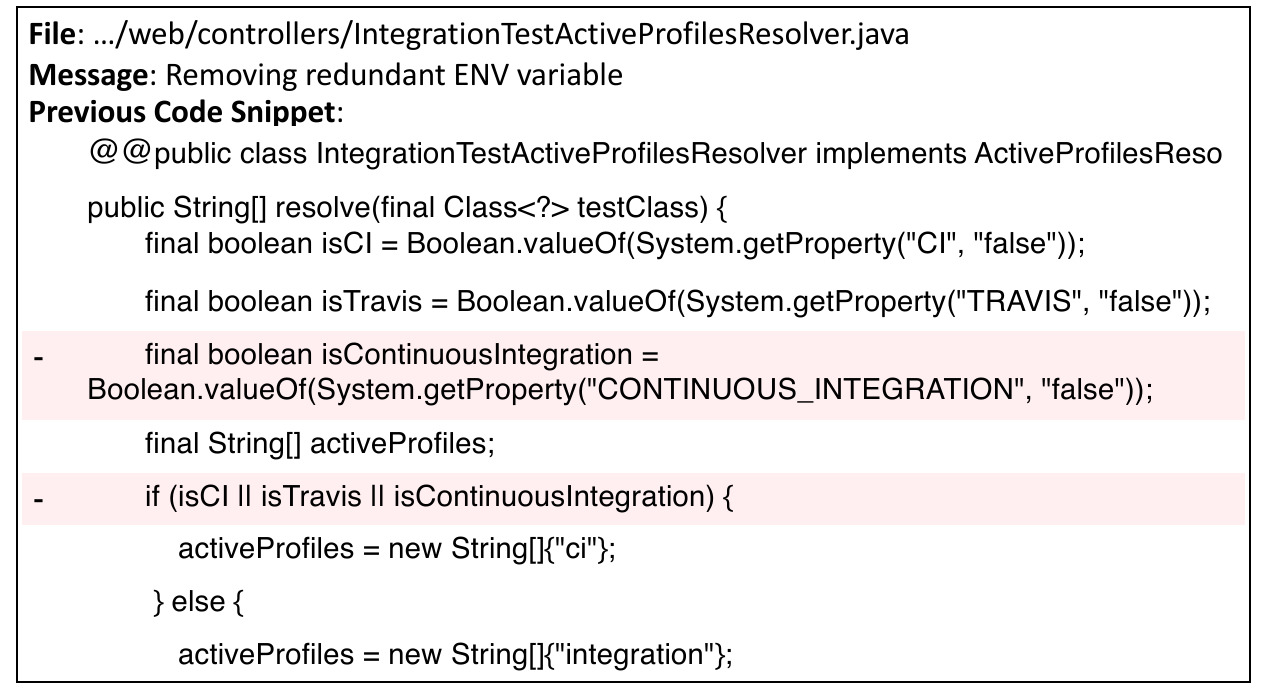}
         \caption{The input for models to generate the updated code snippet.}
         \label{fig:code_gen_source_2}
     \end{subfigure}
     \hfill
     \begin{subfigure}{0.8\textwidth}
         \centering
         \includegraphics[width=\textwidth]{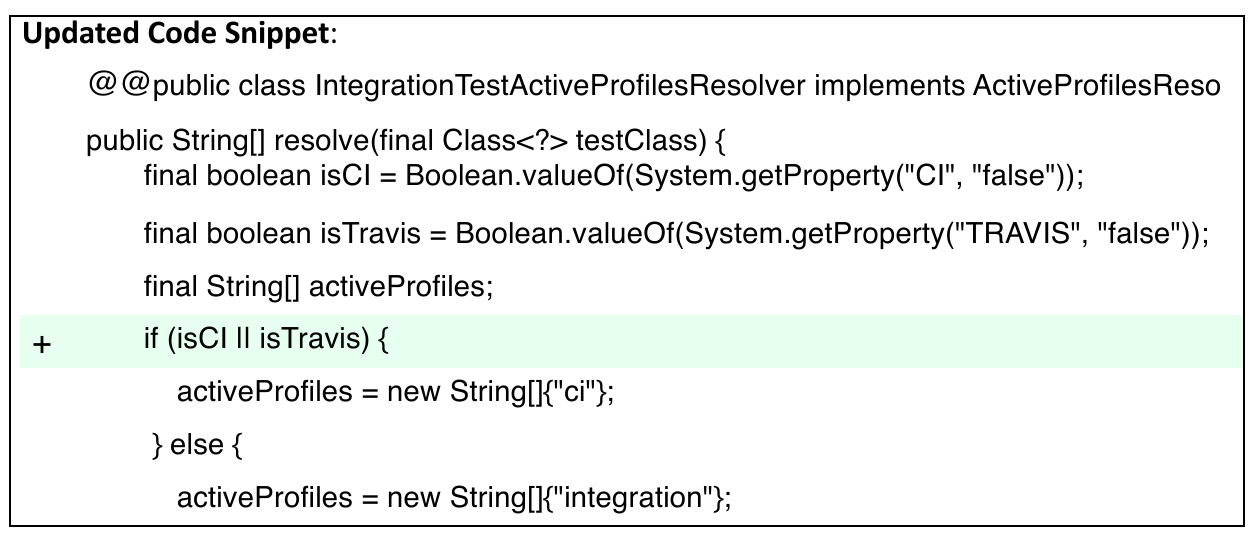}
         \caption{The updated code snippet (i.e., ground-truth).}
         \label{fig:code_gen_ground_2}
     \end{subfigure}
     \begin{subfigure}{0.8\textwidth}
         \centering
         \includegraphics[width=\textwidth]{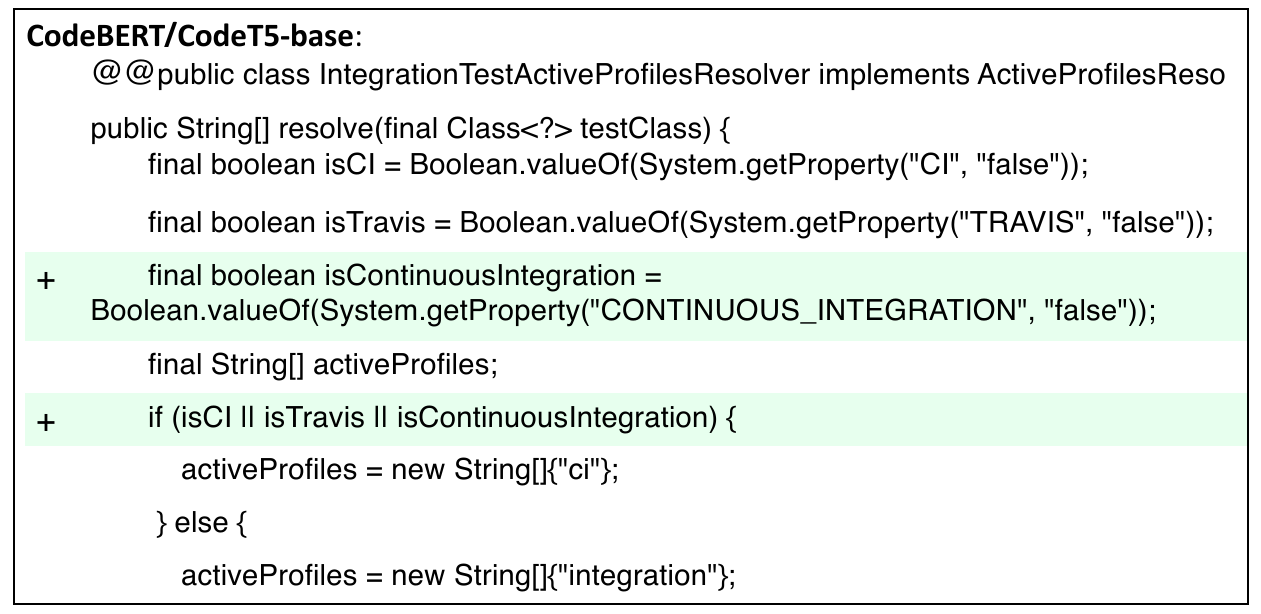}
         \caption{The results provided by CodeBERT and CodeT5-base.}
         \label{fig:code_gen_codebert_2}
     \end{subfigure}
    \caption{Another example from Java for the task of updated code snippet generation where the commit id is \href{https://github.com/Netflix/genie/commit/735d8c29809e997b5ed5978951729b1c33e640a3}{e640a3}.}
        \label{fig:code_gen_2}
\end{figure*}
    \begin{figure*}[ht]\ContinuedFloat
         \centering
     \begin{subfigure}{0.8\textwidth}
         \centering
         \includegraphics[width=\textwidth]{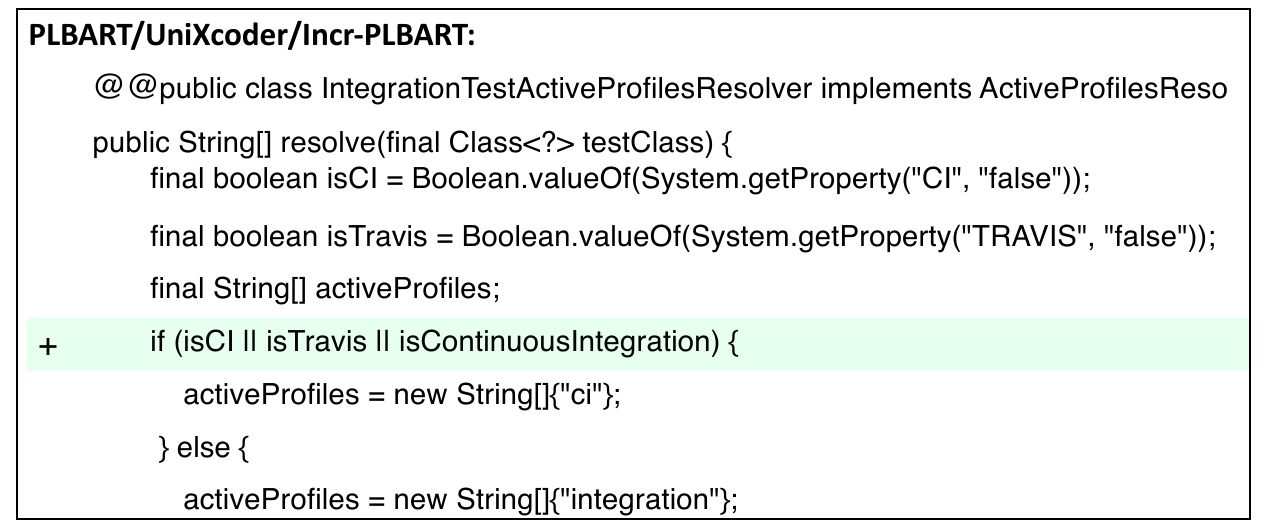}
         \caption{The results provided by PLBART, UniXcoder and Incr-PLBART.}
         \label{fig:code_gen_plbart_2}
     \end{subfigure}
         \hfill
     \begin{subfigure}[b]{0.8\textwidth}
         \centering
         \includegraphics[width=\textwidth]{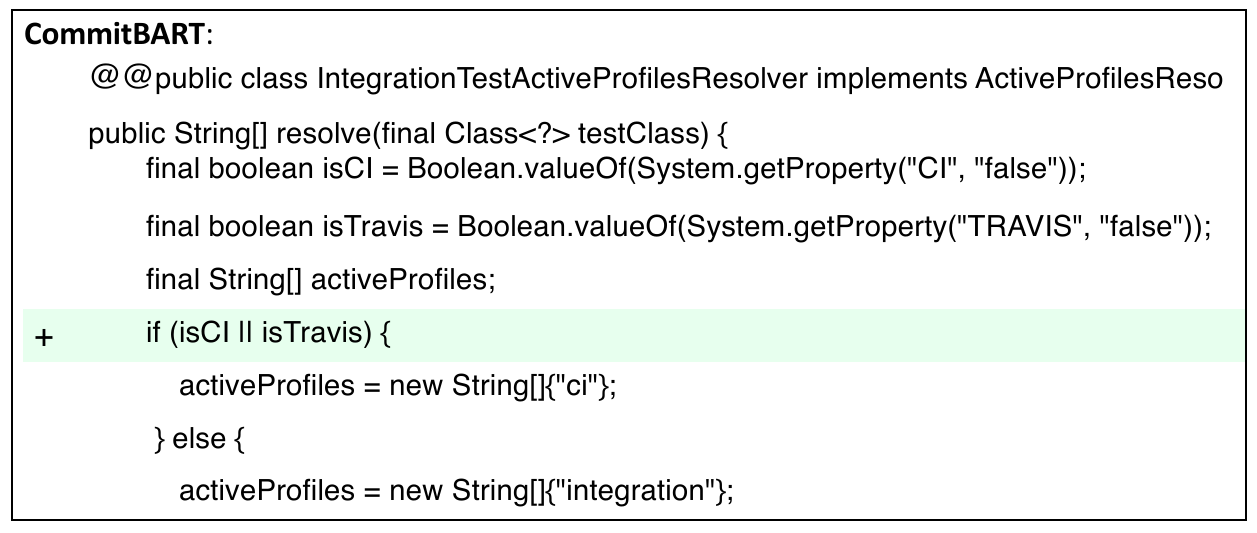}
         \caption{The result provided by CommitBART.}
         \label{fig:code_gen_commitbart_2}
        \end{subfigure}
         \caption{Another example from Java for the task of updated code snippet generation where the commit id is \href{https://github.com/Netflix/genie/commit/735d8c29809e997b5ed5978951729b1c33e640a3}{e640a3} (cont.).}
        \label{fig:code_gen_3}
\end{figure*}

\end{document}